\documentstyle[prl,aps,multicol,epsf,graphicx,xspace,cite,preprint]{revtex}

\newcommand{\critical}[1]{$#1_{\mathrm{c}}$\xspace}
\newcommand{\Tc}{\mbox{\critical{T}}\xspace}

\newcommand{\Ic}{\critical{I}}
\newcommand{\Jc}{\critical{J}}

\newcommand{\etal}{{\sl et\,al.}\xspace}

\newcommand{\ybco}{YBa$_2$Cu$_3$O$_{7-\delta}$\xspace}

\newcommand{\bscco}{Bi$_2$Sr$_2$Ca$_1$Cu$_2$O$_{8+x}$\xspace}

\newcommand{\ycfbco}{Y$_{0.6}$Ca$_{0.4}$Ba$_2$Cu$_3$O$_{7-\delta}$\xspace}

\newcommand{\sto}{SrTiO$_3$\xspace}

\newcommand{\degree}{$^{\circ}$\xspace}

\newcommand{\dxxyy}{d$_{x^2-y^2}$\xspace}

\begin{document}
\draft
{}

\title{\vspace*{1cm}Half\,-\,$h/2 \textrm{e}$ -- Oscillations of SQUIDs}
\date{\today}
\author{C.\,W. Schneider$^1$, G. Hammerl$^1$, G. Logvenov$^1$, T. Kopp$^1$, J.\,R. Kirtley$^2$, P.\,J. Hirschfeld$^3$, and J. Mannhart$^1$}
\address{
$^1$ Experimentalphysik VI, Center for Electronic Correlations and Magnetism, Institute of Physics,
Augsburg University, D-86135
Augsburg, Germany\\
$^2$ IBM Thomas J. Watson Research Center, P.O.Box 218, Yorktown
Heights, New York 10598\\
$^3$ Department of Physics, University of Florida, Gainesville, Florida 32611, USA\\}

\date{\today}
\maketitle \vspace*{0.2cm}
\begin{abstract}
The current-voltage characteristics of Superconducting Quantum Interference Devices (SQUIDs) are
known to modulate  as a function of applied magnetic field with a period of one flux quantum
$\Phi_0=h$/2e. Here we report on the fabrication and properties of SQUIDs modulating with a period
of $1/2\times\Phi_0$.  The characteristics of these bicrystal SQUIDs are consistent with either a
strong sin(2$\varphi$) component of the current-phase relation of the Josephson current, or with an
interaction between the Cooper-pairs, causing an admixture of quartets to the condensate.
\end{abstract}
{~}\\
%


The interference between two superconducting condensates results in a wide range of intriguing, yet
easily accessible phenomena. As shown by B.\,D.~Josephson \cite{Josephson}, interference between
the superconducting order parameters controls the current-voltage ($I$($V$)) characteristic of weak
links. In first approximation, the density of the zero-voltage current traversing such a link is
given  by:
\begin{equation}
J~=~ J_{\mathrm{c}} \sin(\Delta\varphi),
\end{equation}
where \Jc is the critical current density of the junction, and $\Delta\varphi$ describes the
difference of the phases of the superconducting order parameter on both sides of the junction. In
case a voltage is generated by the junction, it is proportional to $\partial_t {\Delta\varphi}:$
\begin{equation}
\partial_t {\Delta\varphi}~=~\textrm{e}^{*}V/\hbar,
\end{equation}
where $\textrm{e}^{*}=2\textrm{e}$ is the charge of the charge carriers in the condensate.
Embedding one or two Josephson junctions into a superconducting loop that encloses an area $A$,
superconducting quantum interference devices (SQUIDs) are obtained \cite{Jaklevic, Clarke}.
Exploiting the phase shifts
\begin{equation}
\vec{\nabla} \varphi~=~(\textrm{e}^{*}/\hbar)\vec{A},
\end{equation}

\noindent induced by the magnetic flux density $\mu_0 \vec H$ penetrating the SQUID loop, these
devices are magnetometers with outstanding sensitivity. Here, $\vec{A}$ is the vector potential
associated with $\vec{H}$. According to Eqs.\,1 and 3, the circulating currents in the SQUIDs, and
thereby the SQUIDs' characteristics, periodically modulate  with $H$, the period $\Delta
\Phi$ being given by the magnetic flux quantum $\Phi_0$
\begin{equation}
\Delta \Phi ~ =~\Phi_0~=~h/\textrm{e}^{*}~=h/2\textrm{e}.
\end{equation}

SQUIDs have been fabricated in huge numbers. In all cases in which the oscillation period was
measured, it was found to equal $h/2\textrm{e}$. In fact,  rings containing grain boundary
Josephson junctions have  been used with great success to measure the Cooper-pair charge in \ybco
\cite{Gough}. Therefore one might believe that the oscillation period of  SQUIDs always has to be
$\Phi_0$. This belief would be erroneous, because neither is $\Delta \Phi=h/2\textrm{e}$ required
for SQUIDs that use junctions for which Eq.\,1 is not fulfilled (see, e.g., \cite{Tanaka}), nor do
Eq.\,1 and 3 predict $\Phi_0$-periods for SQUIDs built from hypothetical superconductors with
carriers
characterized by charges $\mathrm{e}^{*}\neq 2\mathrm{e}$.\\

Intriguingly, experiments measuring the impedance of rf-SQUIDs \cite{Il'ichev} and transport
measurements of dc-SQUIDs \cite{Lindström} indicate the contribution of a $\sin(2\varphi)$
component to the Josephson current for (001)/(110) grain boundaries in \ybco, such that for these
junctions the relation between the Josephson current density and the phase difference is given by
$J=J_{\mathrm{c1}} \sin(\Delta\varphi)-J_{\mathrm{c2}} \sin(2\Delta\varphi)$. Interestingly, for
faceted (100)/(110) boundaries  it was  noted  that the spatially averaged current density may
consist of many harmonics,  even if the local tunnel current density only has a first harmonic
component $J_{\mathrm{c2}}$ \cite{Mintsnew}. In case $J_{\mathrm{c2}}$ dominates $J$, SQUIDs built
from such junctions are
expected to reveal a periodicity $\Delta\Phi=1/2\times\Phi_0$. \\

The second possibility to generate non-$h/2\mathrm{e}$ periods, superconducting carriers with
charges $\mathrm{e}^{*}\neq 2\mathrm{e}$, seems to be implausible. Nevertheless, superconductors
with such carriers can in principle exist. We were led to consider this idea while measuring the
magnetic flux generated by (100)/(110) boundaries in bicrystalline \ybco films \cite{Mannhart}. As
shown by Fig.\,1, in one of these samples the flux is generated in quantities close to $\Phi \simeq
h/4 \mathrm {e} $ and  $\Phi \simeq h/6 \mathrm {e} $. Although these data could not be confirmed
with other samples, we analyzed the possibility that in superconductors such as the cuprates an
interaction between the Cooper-pairs  binds a fraction of  the pairs into particles with charge $
{\mathrm{e}^{*}}=n\times2{\mathrm{e}},~n=2,3,4~...~$.
\\

These considerations suggested the fabrication of SQUIDs from (100)/(110) boundaries to analyze
their oscillation period. The need for such measurements is underlined by measurements of K.~Char
\etal, who found Shapiro steps \cite{Shapiro, Barone} at (100)/(110) boundaries to occur not only
at the standard voltages $V_n=nh/2\textrm{e}f$, but also at voltages $V_n=nh/4\textrm{e}f$
\cite{Char}. Here, $f$ is the microwave frequency used to generate the resonance steps.
Furthermore, studying the critical current \Ic of such a boundary in a \bscco film, D. van
Harlingen and his group noted that the width of the $I_{\mathrm{c}}(H)$ zero-field maximum
had only half  the expected value \cite{CWS}. \\

While it is possible to attribute these experimental results to effects well known in
superconductivity and to peculiar artifacts resulting from inhomogeneities, we saw the need to
analyze the oscillation periods of dc-SQUIDs built from asymmetric 45\degree boundaries to clarify
whether they equal $h/2\textrm{e}$. We note that numerous SQUIDs containing such junctions have
already been fabricated  with the so-called biepitaxial process. To our knowledge a special
oscillation period was not reported for any (see, e.g. \cite{chew92}). This behavior may be the
result of \Jc inhomogeneities masking the true oscillation period. SQUIDs with asymmetric 45\degree
junctions have also been built using the more reproducible bicrystal technology. The dynamics of
these SQUIDs demonstrated the presence of strong second harmonics in the  current-phase relations
of the
junctions \cite{Lindström}.\\

We fabricated and analyzed four bicrystalline dc-SQUIDs containing (100)/(110) boundaries, together
with reference samples for calibration. The experiments were performed with \ybco films (3 samples)
and with a \ycfbco film (1 sample) grown by pulsed laser deposition  to a  thickness of
$\simeq$\,120\,nm. The commercial \sto-substrates contained (100)/(110)-asymmetric grain
boundaries, specified to within 1\degree. The films were patterned into dc SQUID structures by
photolithography and ion-beam etching  before Au contacts were sputter deposited. As reference
samples, nominally identical SQUIDs were grown on symmetric 45\degree bicrystals. The junctions
have widths of $9\,\mu$m, the SQUID holes are symmetric triangles with base lengths of $8\,\mu$m
and heights of $9.5\,\mu$m. A sketch and a micrograph of the devices are given in Fig. 2. All
measurements were conducted in a magnetically shielded room. The SQUIDs had non-hysteretic
current-voltage $I$($V$)-characteristics, their critical currents were measured by tracing the
$I$($V$)-curves, using a voltage criterion of $\simeq 2\,\mu$V. Because the \ycfbco sample behaved
identically to the \ybco SQUIDs, we do not differentiate the two  in the following.\\

The grain boundaries in the bicrystalline \ybco films are faceted, with facet lengths of the order
of $10-100\,$nm. Together with the \dxxyy symmetry of the \ybco, the facets cause anomalous
$I_{\mathrm{c}}(H)$ dependencies \cite{Hilgenkamprmp}. These arise because a large fraction of the
facets act as $\pi$-Josephson
junctions with negative \Ic.\\

Fig.\,3 displays the $I_c$($H$) dependence of a SQUID with an asymmetric 45\degree boundary
measured at 4.2\,K (a), together with the corresponding characteristic of a control SQUID with a
symmetric 45\degree boundary (c). Both curves show the SQUID oscillations superimposed on the
$I_{\mathrm{c}}(H)$ curves of the Josephson junctions. These interferences occur because the width
of the Josephson junctions is comparable to a side length of the SQUID hole. It is noted that for
the SQUID with the asymmetric boundary, \Ic is small at small $H$, which is due to the suppression
of \Ic by the $\pi$-facets. $I_ {\mathrm{c}} (4.2 \, \textrm{K})$ at $H=0$ typically equals $10-100
\, \mu$A for the \ybco samples, which corresponds to $J_ {\mathrm{c}} (4.2 \,
\textrm{K})=10^{3}-10^{4} \, {\textrm{A/cm}}^2 $, typical values for 45\degree boundaries
\cite{Hilgenkamprmp}. In these respects, the SQUIDs behave as expected.
But how large are the oscillation periods?\\

As shown by Fig.\,3, the SQUID oscillations of the control sample have a period of $\simeq2.7 \,
\mu$T. To evaluate the magnetic flux in the loop that corresponds to this applied  field, we
calibrated the flux focussing of the samples. To do so, 24\degree and 36\degree bicrystalline \ybco
films were structured into the sample pattern. Their oscillation periods are $2.6-2.9\,\mu$T
\cite{remark}, very similar to that of the control sample. Because bicrystal SQUIDs with almost
identical geometries have been reported to oscillate with $h/2 {\mathrm{e}}$ \cite{Hans}, it is
evident that the oscillation period of the control SQUIDs is $h/2 {\mathrm{e}}$, corresponding to a
flux focussing factor of 20.2.\\

In contrast to the standard SQUIDs, the oscillation periods of the SQUIDs with the (100)/(110)
boundaries amount to $1.44\, \mu$T (Fig.\,3a,b). Because the flux focussing factor and the loop
areas are nominally identical to those of the control SQUID, this  periodicity equals $\Phi_0/2=h/4
{\mathrm{e}}$. This  period we observed in three out of the four SQUIDs, but never in a control
sample. In the SQUID shown in Fig. 3a, the  period even changes from $h/4 {\mathrm{e}}$ to $h/2
{\mathrm{e}}$, when $H$ exceeds $5\,\mu$T (Fig.\,3). Providing an internal calibration, this curve
gives additional evidence that the oscillation period equals $\Phi_0/2$ at small $H$. Both
oscillation periods are clearly revealed  by the Fourier-transform of the $I_c$($H$)-dependence
(Fig.\,4). The spectra even suggest minute signals at $h/6 {\mathrm{e}}$ and $h/8 {\mathrm{e}}$,
although the higher harmonics die off quickly. The temperature dependencies of both oscillation
periods differ. While at 77\,K the $\Phi_0/2$ have almost completely disappeared,
the $\Phi_0$ oscillations are still  pronounced (see Fig.\,3d).\\

Clearly the dc-properties of the SQUIDs are highly unusual and do not agree with simple
expectations based on the dc-Josephson relation (Eq.\,1). To analyze whether  the rf-properties of
the SQUIDs also differ from the standard behavior as predicted by the ac-Josephson-relation
(Eq.\,2), we measured the $I$($V$) curves of the samples under microwave radiation at 11.88\,GHz.
Since the radiation caused the samples to become noisy at small $H$, these experiments could only
be performed in the  large $H$ regime, for which the samples showed the $h/2 {\mathrm{e}}$ period.
The $I$($V$) characteristics display conventional, integer Shapiro steps $V_n=nh/2\textrm{e}f$ in
case the samples are biased with magnetic fields for which the SQUID oscillation have \Ic-maxima.
Shapiro steps at half the standard frequency $V_n=nh/2\textrm{e}f$ are seen if the junctions are
operated in magnetic fields that result in \Ic-minima.\\

The SQUID properties strongly suggest that they are caused by a dominating
$J_{\mathrm{c2}}$-contribution to the Josephson current, in agreement with Ref.
\cite{Il'ichev,Lindström}. For (100)/(110) boundaries a large $J_{\mathrm{c2}}$/$J_{\mathrm{c1}}$
ratio is expected, because $J_{c1}$ is suppressed by the tunneling geometry, which couples the
nodal with the anti-nodal directions of the order parameters. Further, because $\vec{\nabla}
\varphi$ is insignificant at small $H$ (Eq.\,3), the $\pi$-facets have negative current densities
in this field range: $J_1 = J_{\mathrm{c1}} \sin (\Delta\varphi+\pi) = -J_{\mathrm{c1}} \sin
(\Delta\varphi)$ and therefore suppress $I_{\mathrm{c1}}$. However, these facets do not suppress
$I_{\mathrm{c2}}$, because at the $\pi$-facets $J_2 = J_{\mathrm{c2}} \sin (2\times
(\Delta\varphi+\pi))$. Therefore, the $I_{\mathrm{c2}}/I_{\mathrm{c1}}$ ratio is enhanced at small
$H$, precisely where the $\Phi_0/2$-period is seen. The suppression of $I_{\mathrm{c1}}$ by the
asymmetric 45\degree boundaries also explains why the $h/4 {\mathrm{e}}$-period has only been seen
for SQUIDs fabricated from such junctions. Because the facet structure of the grain boundaries
plays a key role in this mechanism, it is understandable that the fourth sample  we have studied
showed $\Delta \Phi=h/2 {\mathrm{e}}$, as did the SQUIDs fabricated by Lindstr\"om and coworkers.
Due to non-identical growth parameters, the devices
vary in their facet structures and  therefore have different $J_{\mathrm{c2}}(x)$/$J_{\mathrm{c1}}(x)$ ratios.\\

Of course, if the data were interpreted in conventional terms, such that the SQUID oscillations are
taken as a measure of $\textrm{e}^{*}$, the $h/4 {\mathrm{e}}$-oscillations would propose that
carriers with charge 4\textrm{e} are present at the grain boundary, and that their condensate
couples the phases of the two Josephson junctions around the SQUID-hole. The possibility that this
model was correct seems to be remote, because the BCS ground state is composed solely of
Cooper-pairs \cite{Bardeen}. However, the derivation of this ground state is based on the
assumption that all interactions between the charge carriers are two-particle interactions,
described by the term $V_{\vec{k}, \vec{k'} }$. This mean-field approximation is clearly
appropriate for conventional superconductors, in which the coherence length $\xi$ is much larger
than the spacing $d$ between Cooper-pairs. However, because in the cuprates $\xi$ and $d$ are of
the same order, and because correlation effects are predominant, it is not clear why this
approximation should be applicable to the high-\Tc compounds \cite{Kapitulnik,Tschneider}.\\

The possible attractive or repulsive interaction between an electron or a hole of one Cooper-pair
with an electron or a hole of a second  pair, arising for example from magnetic interactions,
Coulomb interactions or lattice distortions, results in  a pair-pair interaction energy
$V_{\mathrm{CP,CP}}$. Depending on the sign of $V_{\mathrm{CP,CP}}$, the two Cooper-pairs are
correlated or anticorrelated. For a strongly attractive $V_{\mathrm{CP,CP}}$ interaction they pair.
For weak $V_{\mathrm{CP,CP}}$, these pairs dynamically rearrange themselves in a fluctuating
manner. The correlations between the pairs result in a non-BCS ground state, expressed by the
Ansatz
\begin{equation}
\Psi ~ =~\alpha_2\Psi_{\mathrm{2e}} + \alpha_4\Psi_{\mathrm{4e}}  +  \alpha_6\Psi_{\mathrm{6e}} + ...,
\end{equation}
where $\alpha_2\Psi_{\mathrm{2e}}$ is the Cooper-pair term and the higher order terms
describe the Cooper-pair multiples, such as quartets or sextets.\\

We note that an analysis of the temperature dependencies of the penetration depths and of the
fluctuation contributions to the conductivity and to the magnetization led to the proposal that the
superconducting phase transition of the cuprates is due to interacting pairs \cite{Tschneider}.
Intriguingly, a related situation exists for the binding of nucleons in light nuclei. While pairs
of nucleons condense into a superconducting condensate, they may bind by their higher order
correlations into multiples, in particular into alpha particle-like quartets \cite{Vollhardt,
Faessler}. The onset of ``quartetting'' in nuclear matter has been considered \cite{Roepke} and
found to predominate in the low
density regime.\\

Because the Josephson current crossing weak links also couples the higher order multiples, SQUIDs
built from superconductors in which $|\alpha_4|^{2}/|\alpha_2|^{2}$ is non-negligible  are expected
to show an $h$/4e contribution in their oscillations. For this to occur, the
$\alpha_4\Psi_{\mathrm{4e}}$ term has to be sufficiently large to provide phase coupling around the
SQUID loop. Although this exotic mechanism is consistent with the measured $h$/4e-behavior of the
SQUIDs, it has to provide answers for the questions why $h$/4e periods have never been observed for
SQUIDs with more conventional junction geometries, and why the Abrikosov vortices carry a flux of
$h$/2e. To account for these effects, one would need to conclude that
$|\alpha_4|^{2}/|\alpha_2|^{2}$ is
enhanced  at (100)/(110) interfaces.\\

One might expect to find non-$h$/2e oscillation periods or non-$h$/2e vortices also in other
systems and under different circumstances, for example half
flux quanta in the bulk, in particular if further mechanisms exist that cause the unusual SQUID properties reported. \\

In summary, we have fabricated three high-\Tc SQUIDs which display  $I_{\mathrm{c}}(H$)
oscillations with a period of $1/2\times h/2\mathrm{e}$. Two mechanisms are found to be consistent
with these characteristics. The more conventional  is based on Josephson currents with a $J =
J_{\mathrm{c}} \sin (2\Delta\varphi)$ current-phase relation. The other mechanism is based on
interacting pairs that form quartets with charge 4e. Further experiments are required to clarify
whether one of these effects underlies
the SQUID behavior, and whether these mechanisms are possibly realized in other  systems.\\

The authors gratefully acknowledge very helpful discussions with  Y.~Barash, B.~Cheska, U.~Eckern,
A.~Faessler, G.~Ingold, V.~K\"orting, C.~Laschinger, R.\,G.~Mints, H.~Rogalla, D.~Vollhardt, and
C.\,C.~Tsuei. This work was supported by the BMBF (EKM-project 13N6918), by the Deutsche
Forschungsgemeinschaft (SFB 484), and by the ESF (PiShift).
%


\begin{figure}[bpth]
\vspace{1cm}
\center\includegraphics[width=0.9\columnwidth,clip=true]{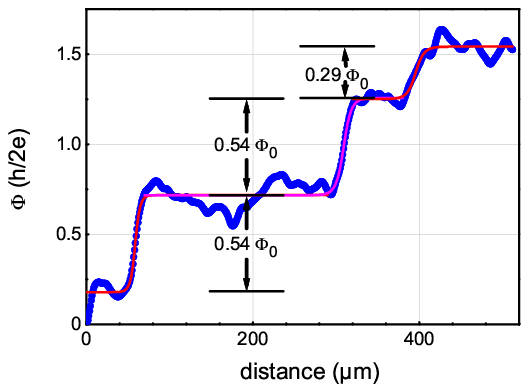}
\vspace{0.5cm}
\caption{\label{fig1}Magnetic flux $\Phi$ generated at 4.2\,K by a (110)/(100) grain boundary in a
\ybco-film, measured by scanning SQUID microscopy.}
\end{figure}

\begin{figure}[bpth]
\vspace{1cm}
\center\includegraphics[width=0.9\columnwidth,clip=true]{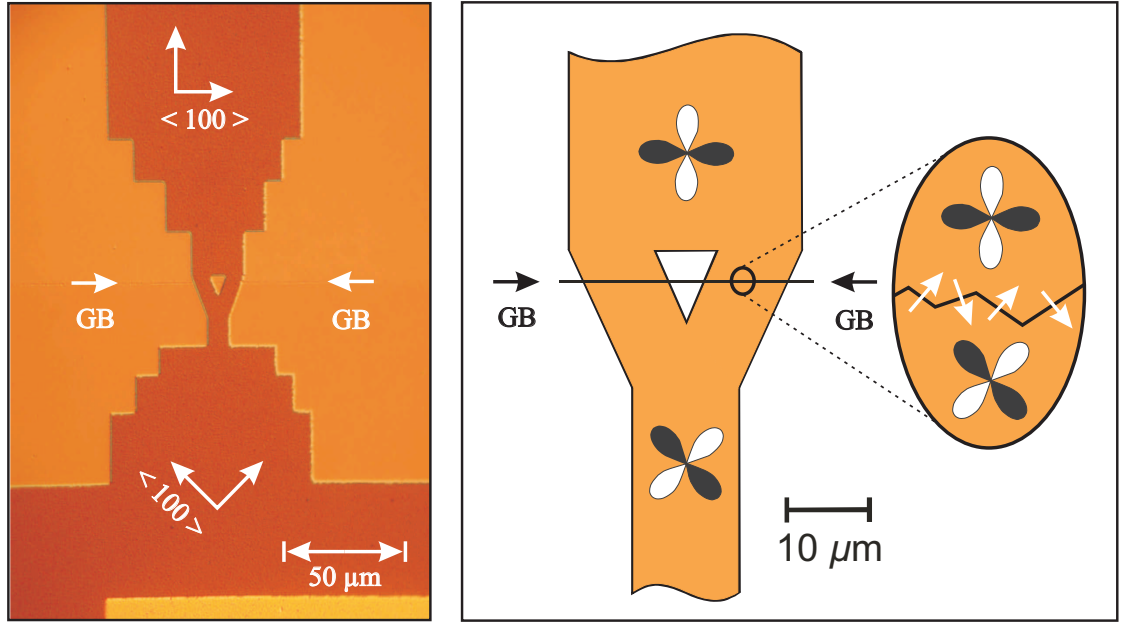}
\vspace{0.5cm}
\caption{\label{fig2}Micrograph (left) and sketch (right) of a \ybco SQUID used in the study. The
grain boundary is marked by the arrows labelled  \lq GB\rq. The blow-up on the right sketches the
facet structure and the corresponding Josephson current crossing the boundary.}
\end{figure}

\begin{figure}[bpth]
\vspace{1cm}
\center\includegraphics[width=0.85\columnwidth,clip=true]{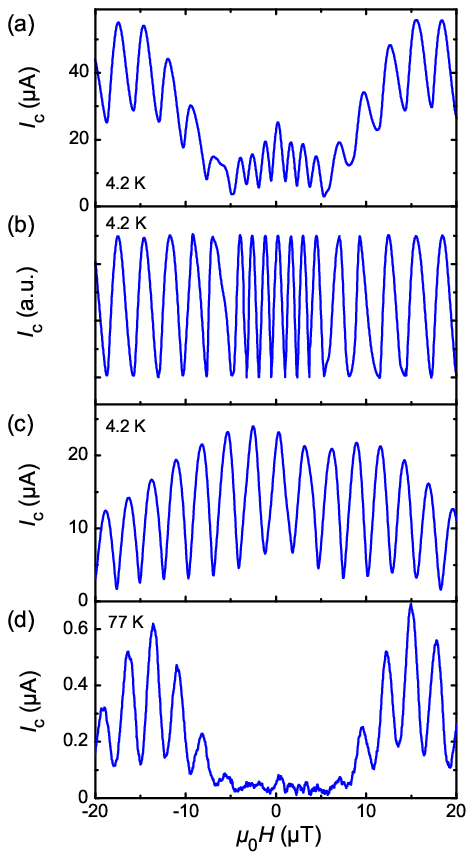}
\vspace{0.5cm}
\caption{\label{fig3} (a) Magnetic field dependence of the
critical current of a \ybco SQUID with an asymmetric 45\degree
grain boundary measured at 4.2\,K; (b) same data as (a) with the
oscillation amplitude numerically normalized; (c) field dependence
of the critical current of a control SQUID with a symmetric
45\degree  boundary; (d) same as (a), but measured at 77\,K.}

\end{figure}

\pagebreak

\begin{figure}[bpth]
\vspace{1cm}
\center\includegraphics[width=0.9\columnwidth,clip=true]{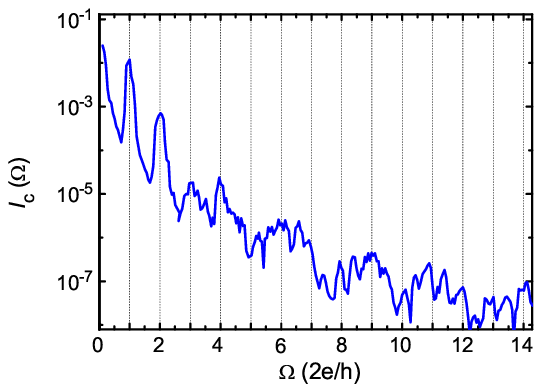}
\vspace{0.5cm}
\caption{\label{fig4} Fourier transform of the SQUID-oscillations shown in Fig. 3(a).\hspace*{4cm}}
\end{figure}


\begin{references}
\bibitem{remark} For the 24\degree SQUID the period was measured at 77\,K, since its \Ic was so high
that stable oscillations could not be obtained at 4.2\,K.
\end{references}
\end{document}